\documentclass[12pt]{iopart}
\usepackage{iopams} %it loads the four AMS packages
\usepackage{graphicx} %for figures
\pdfoutput=1

\begin{document}
%\title[short title]{A phase field crystal model for active particles}
%A phase field crystal model for self-propelled particles PROBABLY BETTER TITLE
\title{A mesoscopic field theoretical approach for active systems}
\author{F. Alaimo$^{1,2}$, S. Praetorius$^1$ and A. Voigt$^{1,2,3}$}
\address{$^1$ Institut f\"{u}r Wissenschaftliches Rechnen, TU Dresden, 01062 Dresden, Germany}
\address{$^2$ Dresden Center for Computational Materials Science (DCMS), TU Dresden, 01062 Dresden, Germany}
\address{$^3$ Center for Systems Biology Dresden (CSBD), Pfotenhauerstr. 108, 01307 Dresden, Germany}	

\ead{francesco.alaimo@tu-dresden.de}

\begin{abstract}
  We introduce a mesocopic modeling approach for active systems. The continuum model allows to consider microscopic details
  as well as emerging macroscopic behavior and can be considered as a minimal continuum model to describe 
  generic properties of active systems with isotropic agents. The model combines aspects from phase field crystal (PFC) models and 
  Toner-Tu models. The results are validated by reproducing results obtained with corresponding agent-based microscopic models.
  We consider binary collisions, collective motion and vortex formation. For larger numbers of particles we analyze the coarsening process
  in active crystals and identify giant number fluctuation in the cluster formation process.
\end{abstract}
%\noindent{\it Keywords\/}: keyword %not required
%\pacs{#1} %pacs number
%\ams{#1} %msc number
\maketitle

\section{Introduction}
%  The first section is normally an introduction, which should state clearly the object
% of the work, its scope and the main advances reported, with brief references to relevant
% results by other workers. In long papers it is helpful to indicate the way in which the
% paper is arranged and the results presented.

  Active systems can exhibit a wide range of collective phenomena. Depending on the density and interaction details, energy taken up on the 
  microscopic scale can be converted into macroscopic, collective motion. Theoretically, such behavior can be addressed
  either from the microscopic scale, taking the interactions into account or from the macroscopic scale, focusing on the emerging 
  phenomena. For reviews on both theoretical descriptions see e.g. \cite{RamaswanyARCMP2010,Cates_RPP_2012,Marchetti2013}. We will here
  introduce a mesoscopic modeling approach which combines aspects from both scales.

  Vicsek-like models \cite{VicseketalPRL1995} are examples for the microscopic viewpoint. These models consider particles, which travel at a 
  constant speed to represent self-propulsion, whose direction changes according to interaction rules which comprise explicit alignment and noise.
  However, explicit alignment rules are not necessary for the emergence of collective phenomena. For elongated particles already the shape gives 
  rise to alignment mechanisms due to steric interactions, see e.g. \cite{PeruanietalPRE2006,GinellietalPRL2010}, and even for circular or 
  spherical particles collective phenomena can be observed if inelastic behavior in the interaction rules is considered, see e.g. \cite{Grossman2008}. 
  Besides these agent-based microscopic approaches 
  also field theoretical macroscopic modeling of active systems has been widely applied. Most approaches are based on the Toner-Tu model \cite{TonerTu1995} and 
  consider only orientational ordering. Collective motion has been addressed within such models, see e.g. \cite{BaskaranetalPRL2008}. More detailed 
  continuum modeling approaches which address besides orientational ordering also positional ordering and thus include also aspects from 
  microscopic models are rare and have so far only been considered for active crystals \cite{Menzel2013,Menzel2014}. Such systems arise for high 
  densities, where particle interactions dominate the propulsion. In this article we propose an extension of the model considered in \cite{Menzel2013} 
  to allow also for low densities. The proposed mesoscopic field theoretical approach can be considered as a minimal continuum model to describe 
  generic properties of active systems with isotropic agents.
  
  After introducing the model, we validate it by reproducing results obtained with corresponding agent-based microscopic models \cite{Grossman2008}. 
  We consider binary collisions, collective motion and vortex formation. Considering larger systems, which can be accessed using the introduced
  mesoscopic field theoretical approach, the formation of collective motion can be analyzed. For high densities we observe a coarsening process of regions 
  of different directions of collective motion. This was already mentioned in \cite{Menzel2013}, but not analyzed. In a broader context the observations can also 
  be related to defects in active crystals. For orientational ordering this was e.g. analyzed in \cite{WeberetalPRL2014}. Another remarkable property of active systems is giant 
  number fluctuations. In contrast to equilibrium systems, where the standard deviation $\Delta N$ in the mean number of particles $N$ scales as 
  $\sqrt{N}$ for $N \to \infty$, in active systems $\Delta N$ can become very large and scales as $N^\alpha$, with $\alpha$ an exponent as large as $1$ in 
  two dimensions \cite{TonerTu1995}. This theoretical prediction is often associated with elongated particles and a broken orientational symmetry   
  \cite{RamaswamyetalEPL2003,ToneretalAP2005,Chateetal_PRL_2006,Narayanetal_Science_2007}, 
  but it has also been verified in simulations of agent-based models for disks with no-alignment rule, see \cite{Fily2012}, and was demonstrated by experiments 
  and simulations in \cite{Palacci2013}. We use large scale simulations to show giant number fluctuations in the proposed mesoscopic field theoretical
  approach.

\section{The model}  
  Our starting point is the active phase field crystal model derived in \cite{Menzel2013,Menzel2014}
  \begin{equation}\label{eq::Menzel}
  \begin{array}{l}
    \partial_t \psi = M_0 \Delta \frac{\delta \mathcal{F}_{\rm pfc}}{\delta \psi} - v_0 \triangledown \cdot \bi{P} \\
    \partial_t \bi{P} = \Delta \left(\alpha_2 \bi{P} + C_2 \bi{P}^3 \right) - \left( \alpha_4 \bi{P} + C_4 \bi{P}^3 \right) - v_0 \bi{\triangledown} \psi.
  \end{array}
  \end{equation}
  It combines the phase field crystal (PFC) model, introduced in \cite{Elder2002,Elder2004} to model elasticity in crystalline materials, with 
  a simplified Toner-Tu model (without the non-linear convective terms) \cite{TonerTu1995} for active systems. The energy functional 
  $\mathcal{F}_{\rm pfc}$ the PFC model is based on, is a Swift-Hohenberg energy \cite{SwiftHohenberg1977}
  \begin{equation}\label{EQ::PFC_energy}
    \mathcal{F}_{\rm pfc} = \int \left[ \frac{\psi^4}{4} + \frac{1}{2} \psi (r + (1 + \Delta)^2)\psi \right] d \bi{r},
  \end{equation}
  for a one-particle density field $\psi(\bi{r}, t)$, which is defined with respect to a reference density $\bar{\psi}$. The parameter $r$ is related 
  to an undercooling and together with $\bar{\psi}$ determines the phase diagram. The functional arises by splitting the Helmholtz free energy 
  in an ideal gas contribution and an excess free energy, rescaling and shifting $\psi$, expanding the ideal gas contribution in real-space and the excess 
  free energy in Fourier-space, and simplifications by removing constant and linear terms that would vanish in the dynamical equations. A detailed 
  derivation of the energy and its relation to classical density functional theory can be found in \cite{Elderetal_PRB_2007,vanTeeffelenetal_PRE_2009}. 
  The second quantity in eq. (\ref{eq::Menzel}) is the polar order parameter $\bi{P}(\bi{r}, t)$ and the remaining parameters are: $M_0$ mobility, $v_0$ 
  self-propulsion determining the strength of the activity, $\alpha_2$ and $\alpha_4$ two parameters related to relaxation and orientation of the polarization 
  field, and $C_2 $ and $C_4$ are parameters which govern the local orientational ordering. The model is used in \cite{Menzel2013,Menzel2014} to study 
  crystallization in active systems. To allow for 
  a description of individual particles, we consider a variant of the PFC model, the vacancy PFC (VPFC) model, introduced in \cite{Chan2009,Berry2011} and also 
  considered in \cite{Praetorius2015}. Instead of the Swift-Hohenberg functional eq. (\ref{EQ::PFC_energy}) we consider a density field 
  with positive density deviation $\psi$ only, which leads to a modification of the particle-interaction and allows to handle single particles, as well 
  as many individual particles. The new energy functional $\mathcal{F}_{\rm vpfc}$ is:
  \begin{equation}\label{EQ::VPFC_energy}
    \mathcal{F}_{\rm vpfc} = \mathcal{F}_{\rm pfc} + \int H(|\psi|^3 - \psi^3) d \bi{r},
  \end{equation}
  with a penalization parameter $H$. As in other, more coarse-grained models for active systems  \cite{Marchetti2013} we use
  a classical transport term with advection velocity $v_0 \bi{P}$ for the local density field $\psi$. This modification turns out to be more stable in comparison 
  to the term used in eq. (\ref{eq::Menzel}), if considered for individual particles. The second modification ensures the polar order parameter $\bi{P}$ 
  to be a local quantity that is different from zero only inside the particles. The new set of dynamical equations we obtain is:
  \begin{equation}\label{eq::dynamic}
  \begin{array}{l}
    \partial_t \psi = M_0\Delta \frac{\delta \mathcal{F}_{\rm vpfc}}{\delta \psi} - v_0 \triangledown \cdot (\psi \bi{P}) \\
    \partial_t \bi{P} = \Delta \left(\alpha_2 \bi{P} + C_2 \bi{P}^3 \right) - \left( \alpha_4 \bi{P} + C_4 \bi{P}^3 \right) 
    - v_0 \bi{\triangledown} \psi - \beta \bi{P} 1_{\psi \leq 0},
  \end{array}
  \end{equation}
  with $\beta$ a parameter, which is typically larger than the other terms entering the $\bi{P}$ equation.
   
\section{Model verification}  
  The aim of this Section is to show that the mesoscopic field theoretical model eq. \eref{eq::dynamic} can be used to simulate active particles and allows to recover known
  phenomena.  The versatility of the model thereby allows us to apply it to different physical situations that have been previously studied using agent-based models, see e.g. 
  \cite{Grossman2008,Lushi2013}. We first consider the situation 
  of one particle, followed by studying the interaction of two particles. These simple computations allow for detailed parameter studies. We found no qualitative
  difference in the results of our simulations when the parameters $C_2$ and $C_4$ are set to zero. Therefore, we simplify our model by restricting ourselves to 
  the case $C_2 = C_4 = 0$, which allows only gradients in the density field $\psi$ to induce local polar order. The remaining parameters to specify are 
  $M_0$, $v_0$, $\alpha_2$, $\alpha_4$, $\beta$, $H$ and $r$. The second test concerns the 
  emergence of collective phenomena in confined geometries with systems with a number of particles $\simeq 100$.
  
  All the simulations are in two dimensions. The maxima in the local one-particle density field $\psi(\bi{r}, t)$ are always tracked for post-processing and 
  evaluation, and every maximum is interpreted as a particle (see \cite{Robbins2012} for a discussion about the validity of this interpretation). The particle's velocity 
  is computed as the discrete time derivative of two successive maxima.
  
\subsection{Onset of movement and particle shape}
  We at first want to understand what happens in a minimal system, where a single active particle is free to move.
  In particular, we are interested to know if there is a critical value for the activity $v_0$ required for the onset of movement, 
  as it has been observed in \cite{Menzel2013} for active crystals. 
  
\begin{figure}
  \includegraphics{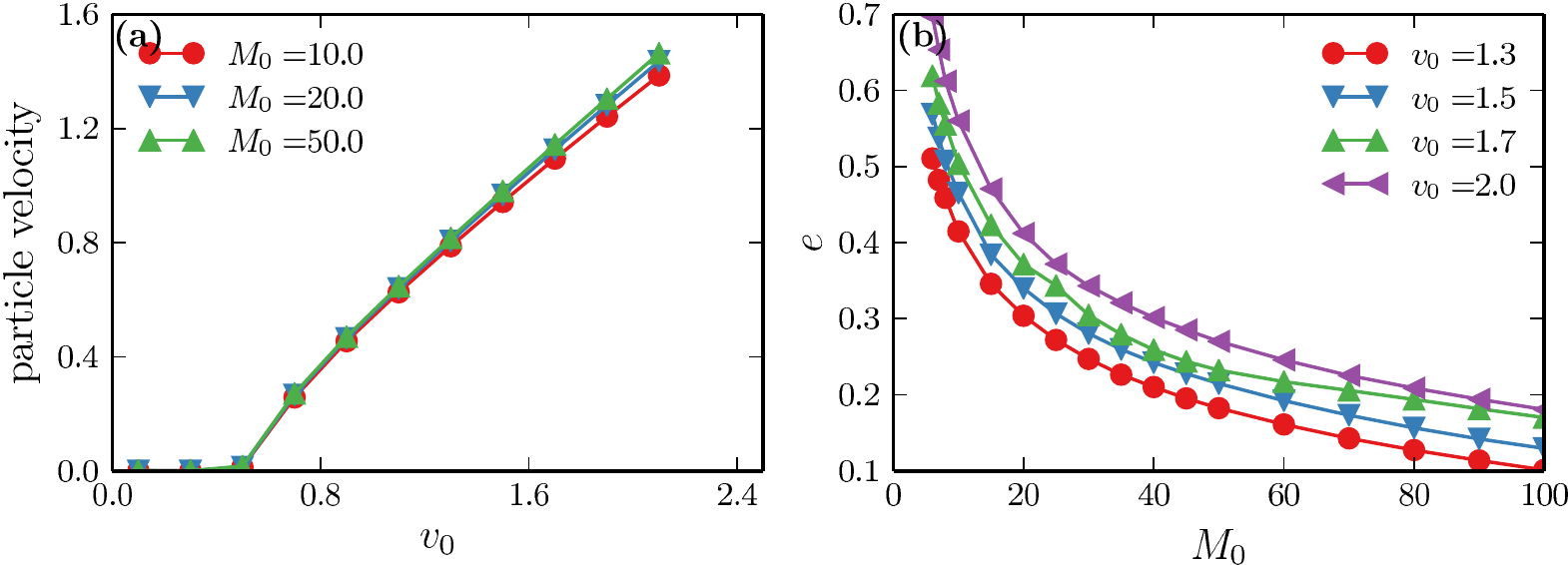}
  \caption{(a) Velocity of a particle as a function of the activity strength $v_0$ for different values
  of the mobility $M_0$. At a threshold value $v_0 \simeq 0.5$ 
  the particle starts to move. Other parameters are $(\alpha_2, \alpha_4, \beta, H, r) = (0.2, 0.1, 2, 1500, -0.9)$ also for the
  other figures unless otherwise specified.
  (b) Eccentricity $e$ of a single particle as a function of mobility. The eccentricity 
  is defined as $e = \sqrt{1-b^2/a^2}$ where $a,b$ are the length of the semi-major and semi-minor axis,
  respectively. For small values of 
  $M_0$ the particles have the form of an elongated ellipse, whereas for larger $M_0$ their form is similar to a circle.}
  \label{figure1}
\end{figure}

  In Fig. \ref{figure1}(a) the particle velocity is plotted as a function of the activity $v_0$ and we can see that for small 
  activities ($v_0 < 0.5$) 
  the particle does not move at all. After a certain threshold value $v_{0} \simeq 0.5$ the particle starts to move with a constant velocity, which 
  approximately linearly increases for increasing $v_0$. This is exactly the same behavior as observed in \cite{Menzel2013} for the 
  sample-averaged magnitude of the crystal peak velocities.

  An important new feature of our model has to do with the mobility term $M_0$ entering eq. \eref{eq::dynamic}. $M_0 = 1$ is used in \cite{Menzel2013}. This value would lead 
  to strong numerical instabilities for the modified model eq. (\ref{eq::dynamic}). Larger
  values for $M_0$ can suppress this numerical instability. While the mobility does not particularly change the particle velocity, we observe that
  $M_0$ directly influences the shape of the particle: for small (but still greater than $1$) 
  $M_0$ the particle shows an elliptic form, whereas further increasing $M_0$ restores a circular shape for the particle. The dependency of the particle shape on the value of $M_0$ is
  shown in Fig. \ref{figure1}(b) for different values of $v_0$ above the threshold value. 
  
\subsection{Binary collisions and elastic deformation}\label{sec::BinaryCollision}
\begin{figure}
  \centering
  \includegraphics{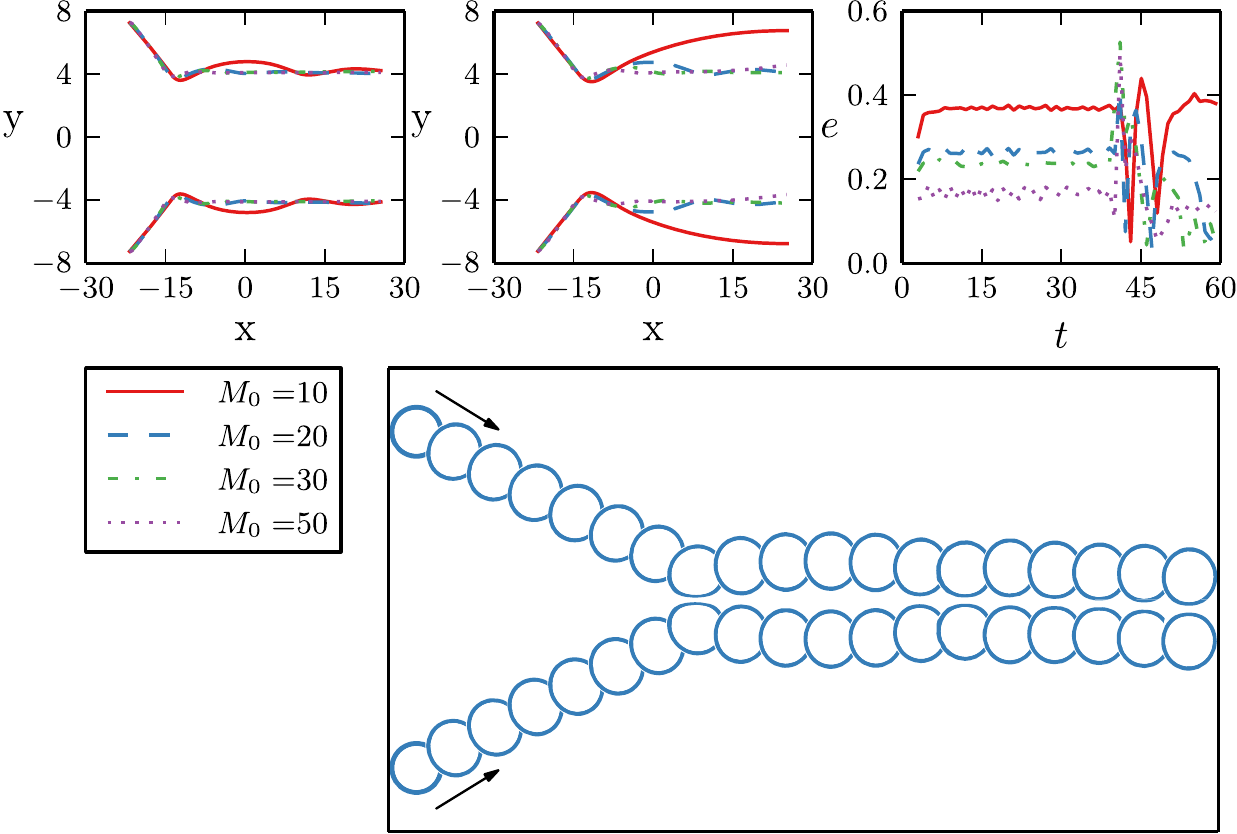} 
  \caption{(a)-(b) Two particles colliding in a perfectly symmetric way for (a) $v_0 = 1.5$ and (b) $v_0 = 2.5$.
  The net effect of the collision is an alignment of the particles direction.
  (c)Eccentricity of a particle as a function of time during a collision  for $v_0 = 1.5$ and different mobility values. We observe a sudden change
  in the particle eccentricity at $t \simeq 40$, corresponding to the collision time.
  (d) Time series of a two-particle collision for $v_0 = 1.5$ and $M_0 = 10$. Notice how the form
  of the particles is slightly changed during the collision. The shape of the particles is identified as a fixed contour line of $\psi$.}
  \label{figure2}
\end{figure}

  The study of binary collisions between particles is often used as a benchmark problem to predict how larger systems evolve, see e.g. \cite{Lushi2013,Lober2015}.
  In particular it has been observed \cite{Grossman2008} that completely inelastic collisions
  lead to a force that aligns the particles direction. We here consider only perfectly symmetric collisions meaning that the incidence angle and the initial velocity are the same for both particles.
  Different particle trajectories obtained using different mobility $M_0$ and activity $v_0$ are shown in Fig. \ref{figure2}(a) and (b). 
  The elastic deformation of a single particle during a collision can be seen in Fig. \ref{figure2}(c), where the eccentricity is plotted 
  as a function of time, whereas a time series of a single collision is shown in Fig. \ref{figure2}(d). Collisions of deformable particles have also been considered in agent-based 
  models \cite{Menzel2012}, with a qualitatively similar behavior. However, the deformations in our approach strongly depend on $M_0$ and are negligible for 
  large values. We therefore do not analyze this effect further and interpret the particles as being spherical in the coming simulations, which are all done for large
  $M_0$.
  
  All results indicate the particle alignment to be not instantaneous. There is an initial oscillatory phase, whose length and magnitude depends on $M_0$
  and $v_0$. Small activity and large mobility lead to an almost instantaneous alignment, whereas large activity and small mobility lead to oscillations for a certain
  period of time, before the particles finally align and travel together.
  
\subsection{Collective motion in confined geometries}
  As already analyzed using agent-based, e.g. \cite{Grossman2008,Menzel2012}, and phase field models \cite{Lober2015,Marth2016} the interaction of a 
  moderate number of particles can lead to collective motion. We here consider simulations with $\simeq 100$ particles to recover these results. To analyze 
  the phenomena we define the translational order parameter $\phi_T$ 
  and the rotational order parameter $\phi_R$ as
  \begin{eqnarray}\label{EQ::orderParameter}
    &\phi_T(t) = \frac{1}{n} \left| \sum_{i=1}^n \hat{\bi{v}}_i(t) \right|, \qquad \phi_R(t) = \frac{1}{n} \sum_{i=1}^n \hat{\bi{e}}_{\theta_i(t)} \cdot \hat{\bi{v}}_i(t)
  \end{eqnarray}
  where $\hat{\bi{v}}_i(t)$ is the unit velocity vector of particle $i$ at time $t$, $\hat{\bi{e}}_{\theta_i(t)} = 
  (-\sin(\theta_i(t)), \cos(\theta_i(t)))$ is the unit angular direction vector of particle $i$ at time $t$, and $n$ is the number of particles. In case of translational migration we obtain $\phi_T = 1$ and 
  rotational migration leads to $\phi_R = \pm 1$.
  
\subsubsection{Collective migration}
\begin{figure}
  \centering
  \includegraphics{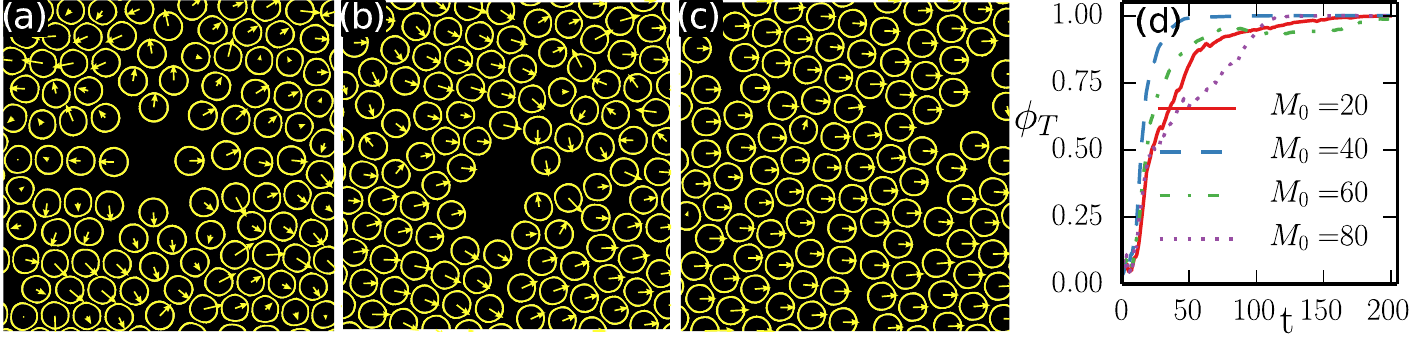}  
  \caption{(a)-(c) Snapshots of a single simulation of $n \simeq 100$ active particles in a square with periodic 
  boundary conditions for $v_0 = 1.5$ and $M_0 = 50$. After an initial chaotic phase particles travel together in the same direction. A movie
  for the evolution is provided in the supplementary material.
  (d) Translational order parameter $\phi_T$ for different values of $M_0$: mobility does not seem to affect the 
  emergence of collective migration. Each curve has been obtained as the average of $10$ different simulations
  started with different initial conditions and $v_0 = 1.5$.}
  \label{figure3}
  \end{figure}

  We consider a square domain with periodic boundary conditions, i.e. simulating an infinite plane where particles
  are free to move without obstacles. Fig. \ref{figure3}  shows the resulting behavior: At the beginning (Fig. \ref{figure3}(a)) 
  there is no specific order and particles move towards different directions. After the first collisions take place some particles start
  to align with each other and small blocks of particles, in which particles orient in the same direction are formed (Fig. \ref{figure3}(b)). 
  If these blocks collide they change their direction until all particles in the system are traveling in the same direction (Fig. \ref{figure3}(c)). 
  This behavior is further confirmed by the translational order parameter $\phi_T \simeq 1$ after a certain time, see Fig. \ref{figure3}(d). 
  We also observe that this migration state is not particularly affected by the value of the mobility $M_0$.
  
\subsubsection{Vortex formation and oscillatory motion}
\begin{figure}
 \centering
  \includegraphics{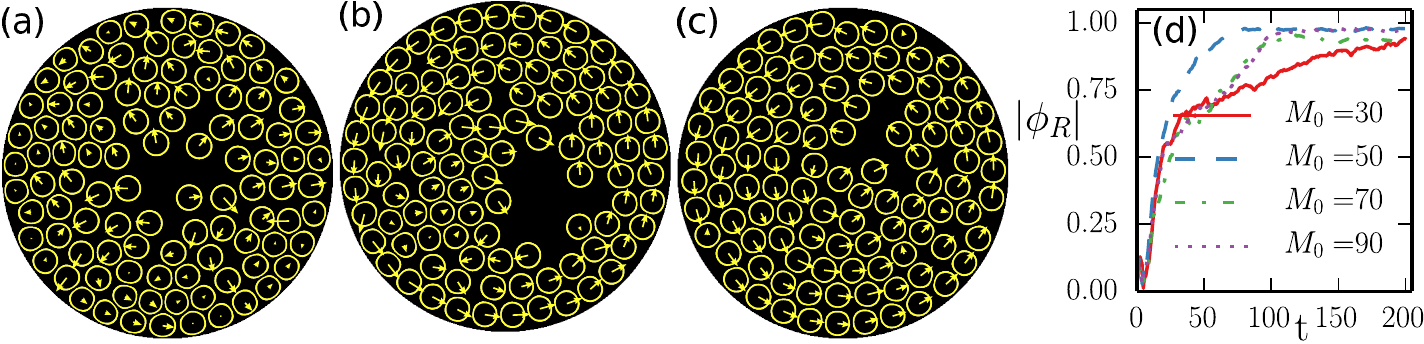} 
  \caption{(a)-(c) Snapshots of a vortex formation by active particles confined in a disk for $v_0 = 1.5$ and $M_0 = 60$. A movie
  for the evolution is provided in the supplementary material.
  (d) The rotational order parameter $| \phi_R |$ shows that after a transient phase the particles
  follow a circular motion for different values of $M_0$. Also in this case, each curve has been obtained as the 
  average of $10$ different simulations
  started with different initial conditions and $v_0 = 1.5$.}
  \label{figure4}
  \end{figure}

  We now consider a confined geometry and specify $\psi = 0$ and $\bi{P} = 0$ at the boundary, which serve as an approximation for reflecting boundary conditions.
  The first geometry is a disk. Here again at the beginning the particles move in a chaotic way, see Fig. \ref{figure4}(a). Then a vortex is formed and most of the
  particles follow an anticlockwise trajectory. In the center some particles move in the opposite direction, see Fig. \ref{figure4}(b).  Eventually also these particles are
  forced to align with the rest of the system, see Fig. \ref{figure4}(c). This behavior is confirmed by the rotational order parameter $\phi_R$, which is shown as a
  function of time in Fig. \ref{figure4}(d). Similarly to the collective migration case studied above, the mobility $M_0$ does not play a major role in the formation of
  the vortex.
 
\begin{figure}
\centering
  \includegraphics{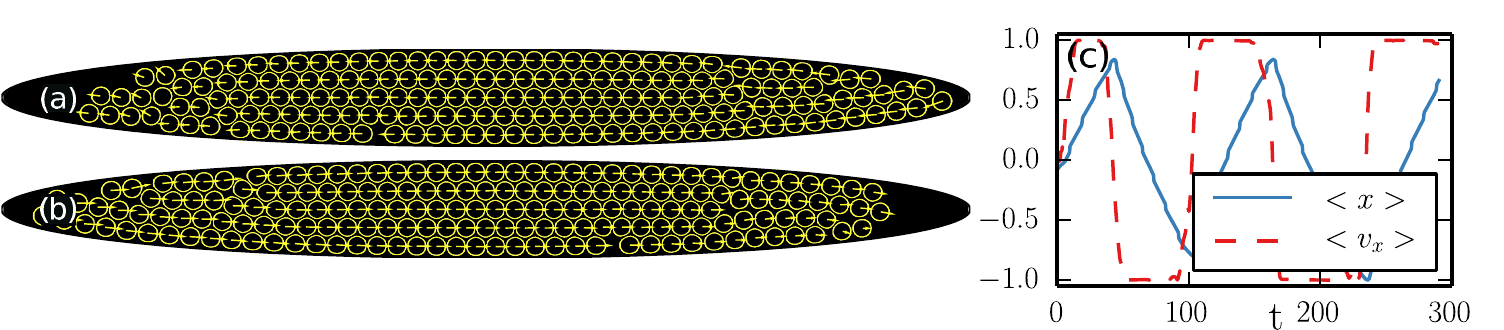}
  \caption{(a)-(b) Snapshots of two different moments of the collective travel of active particles inside an elongated ellipse with $A/B = 10$. Particles
  move together along the major axis and change orientation when they reach a boundary. A movie
  for the evolution is provided in the supplementary material.
  (c) There is an oscillating behavior along the $x$ direction. Other simulation
 parameters are $(v_0, M_0) = (1.5, 500)$}
  \label{figure6}
\end{figure}

  Ellipses provide a more interesting geometry and confining active particles inside them can give rise to different kind
  of collective phenomena, where the ellipse aspect ratio $A/B$, where $A,B$ are the length of the semi-major and semi-minor axis,
  respectively, plays an important role. An ellipse with a small $A/B \leq 3$ show a similar behavior as the disc shape.
  Such geometries produce once again a vortex, where the particles move along the boundaries. More elongated shapes with $A/B = 10$ 
  dramatically change the behavior. As already shown in \cite{Grossman2008} particles move collectively along the major axis with oscillating direction. The same
  behavior could be observed with our model, see Fig. \ref{figure6}. All particles move in one direction until they hit the high curvature region. This produces an 
  impulse that propagates fast along the whole system and reverts the direction of the particles. The whole process repeats every time a boundary is reached and an  
  oscillatory motion is the result. To obtain this result and ensure a constant particle number it is necessary to use a particularly high value for the mobility,
  $M_0 = 500$. 
 
\subsubsection{Validation and numerical issues}  
  These examples demonstrate the validity of our continuous modeling approach. All known qualitative properties which have been shown using agent-based  
  simulations could be reproduced. Until now a sequential finite element approach has been used to solve the evolution equation. For larger
  systems we must work in a parallel environment with multiple processors. We adopt a block-Jacobi preconditioner \cite{Chao2013,PraetoriusDefense2015} that
  allows us to use a direct solver locally. The approach is implemented in AMDiS \cite{Vey2007,Witkowski2015} and shows good scaling properties, which allows to 
  consider systems with $\simeq 15,000$ particles on the available hardware. 
  
\section{Results}
\begin{figure}
\centering
 \includegraphics{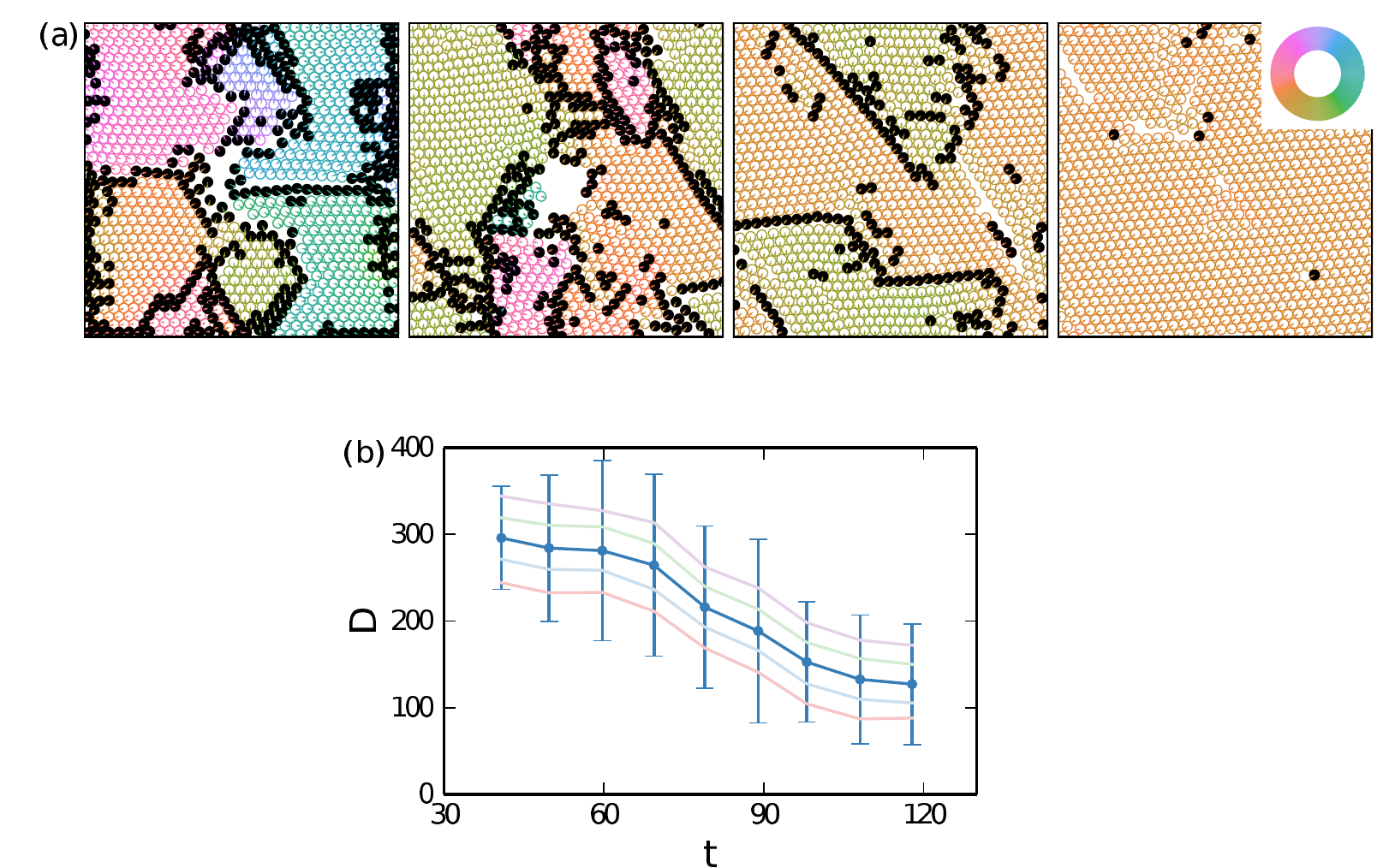}
 \caption{(a) Snapshots of $\simeq 1,000$ particles inside a square with periodic boundary conditions.
 Different colors correspond to different orientations, particles colored in black are those where there is a
 change in the orientation. (b) Decrease of the number of orientational defects $D$ as a function of time. 
 Average and standard deviation of the data for 6 different simulations started with different initial conditions 
 and a tolerance parameter equal to $\pi/10$ are shown in blue. Each of the shaded curves has been obtained 
 as the average of the six simulations, but using different values for the tolerance parameter, ranging from
 $\pi/8$ (purple curve) to $\pi/12$ (red curve). Simulation parameters are $(v_0, M_0) = (2, 60)$}
 \label{figure7}
\end{figure}

  We again consider a system with periodic boundary conditions for which collective migration and cluster formation is expected also for larger numbers of particles. 
  However, how the collective migration state is reached is not well understood and will be analyzed in detail. We start with a system with high volume fraction. 
  The volume fraction is defined as $\phi = n \sigma / |\Omega|$, with number of particles $n$, domain size $|\Omega$ and $\sigma$ 
  the area occupied by a single particle, which is equal to $\sigma = \pi (d/2)^2$, with $d = 4 \pi/\sqrt{3}$ the lattice constant determined by the free energy eq. 
  (\ref{EQ::PFC_energy}). For $\phi > 0.6$ the system shows behavior of active crystals. Figure \ref{figure7}(a) shows various snapshots of the evolution with 
  regions of particles moving in the same direction color coded. The regions can be identified as active grains, which undergo a coarsening process. The black
  particles determine orientational defects. They are identified as particles where the change in orientation from one particle to its neighbors is above a certain 
  threshold. The number of black particles certainly depends on the threshold value, however its decrease is independent on the value. The data is not sufficient to 
  identify a scaling law. However, the robustness of the coarsening process is shown in Fig. \ref{figure7}(b).
 
\begin{figure}
 \includegraphics{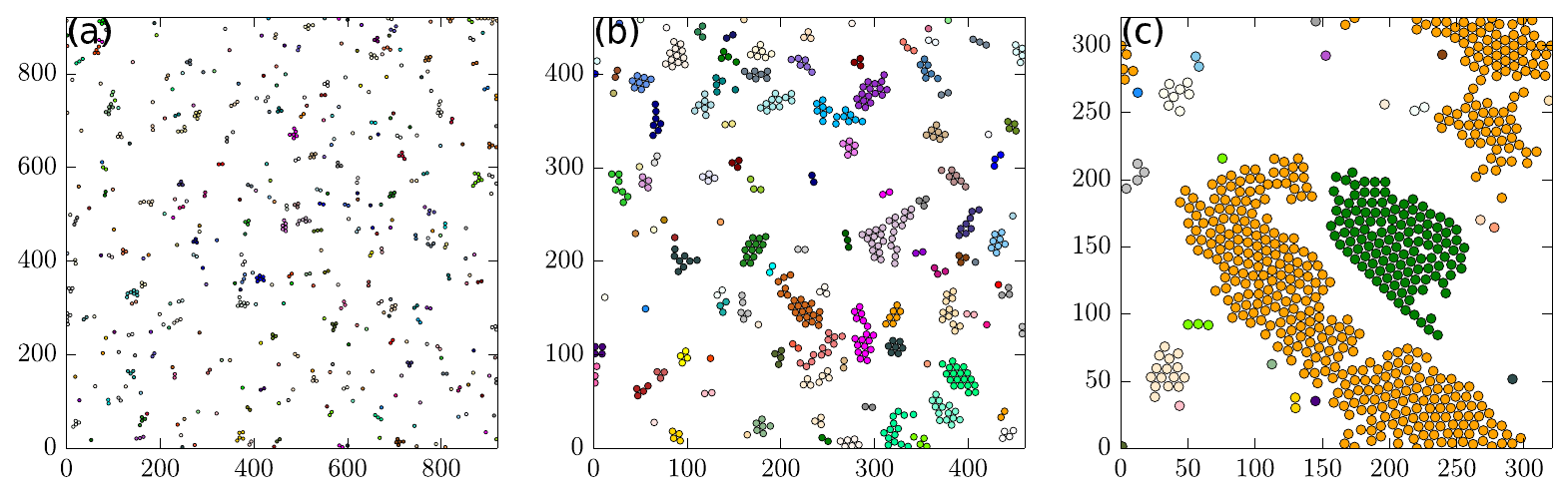}
 \caption{Snapshots of systems having different particle density $\phi$. Particles with the same color
 belong to the same cluster. (a) For $\phi = 0.03$ no cluster is present.
 (b) $\phi$ is increased until $0.12$ and some bigger clusters appear. (c) We clearly observe two big clusters
 when $\phi = 0.25$. Other simulation
 parameters are $(v_0, M_0) = (1.5, 50)$. A movie for the evolution for $\phi = .25$ is provided in the supplementary material. Here the color coding of each particle
 corresponds to the cluster it belongs to initially, which highlights to dynamics during cluster formation.  
}
 \label{figure8}
\end{figure}

  If we decrease the volume fraction the behavior changes. For very low density, $\phi = 0.03$, Fig. \ref{figure8}(a) shows the tendency of the particles to group 
  together, but the formed clusters are very small and there remain many isolated particles in the system. Increasing the density, $\phi = 0.12$, increases the size 
  of the formed clusters, but the average size of a cluster remains very small if compared to the total number of particles, see Fig. \ref{figure8}(b). Further increasing 
  the density, $\phi = 0.25$, leads to the formation of large mobile clusters and a drastic reduction of the number of particles which do not belong to any cluster, see 
  Fig. \ref{figure8}(c). This behavior is similar to the results in \cite{ButtinonietalPRL2013} for (quasi-) two-dimensional colloidal suspensions of self-propelled particles. 
  For these systems we compute the standard deviation $\Delta N$ as a function of the mean number of particles $N$. For active systems it is theoretically predicted 
  that $\Delta N$ scale as $N^\alpha$, with $0.5 < \alpha \leq 1$ and giant number fluctuations occur if $\alpha \approx1$, see \cite{TonerTu1995}. 
  
\begin{figure}
\centering
\includegraphics{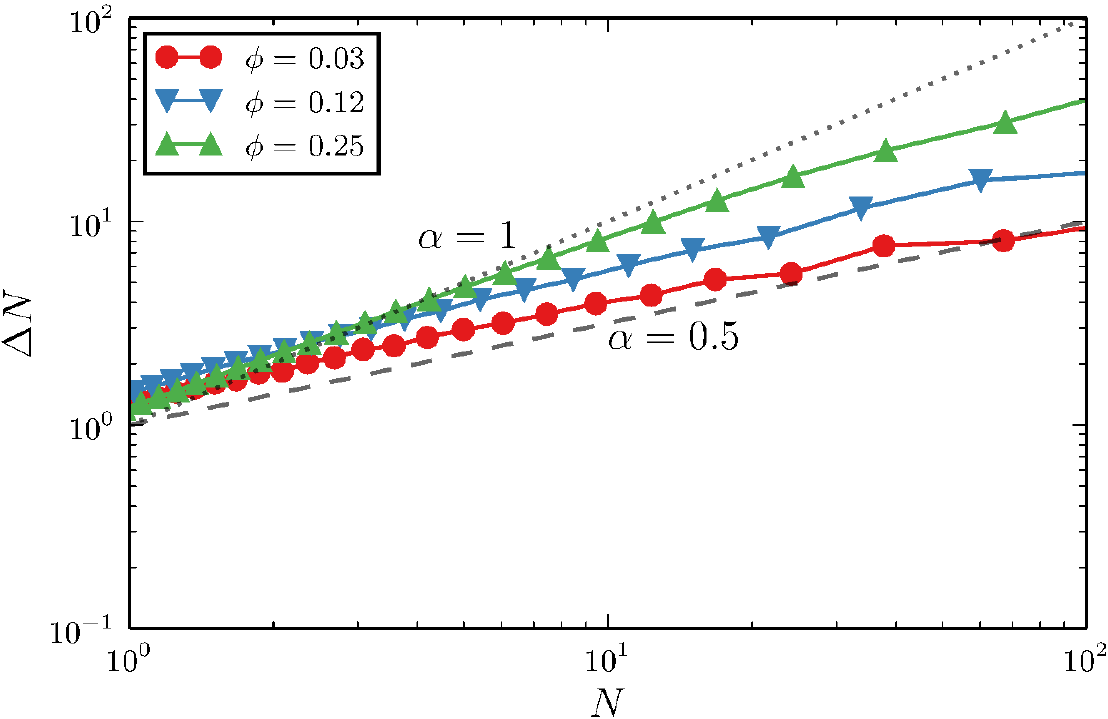}
 \caption{Number fluctuations for the three different values of $\phi$ shown in figure \ref{figure8}. The dashed and dotted
 line corresponds to the case $\alpha = 0.5$ and $\alpha = 1$, respectively.}
 \label{figure9}
\end{figure}

  We compute $\alpha$ by considering different
  subregions of our computational domain. The results for $\simeq 600$ particles are shown in Fig. \ref{figure9}, demonstrating an increase of $\alpha$ with 
  increasing $\phi$, with the largest value reached being $\alpha = 0.79$. This continuous increase in $\alpha$, as well as the obtained values are consistent
  with the behavior found in \cite{Fily2012} for moderate numbers of particles but larger volume fractions. However, there is a significant difference, the formed 
  clusters in \cite{Fily2012} are stationary, our clusters are mobile, similar to the light activated living crystals in \cite{Palacci2013}. The experimental results as well as 
  the simulations in \cite{Palacci2013} lead to similar values of $\alpha$ as ours already for smaller $\phi$. 
  
\section{Summary}

In summary, we extended the phase field crystal model for active crystals \cite{Menzel2013,Menzel2014}, which combines the classical phase field crystal model of Elder et al. \cite{Elder2002,Elder2004} with the approach of Toner and Tu \cite{TonerTu1995}, by considering individual active particles. This can be realized by penalizing a locally vanishing one-particle density, as considered in \cite{Chan2009}. The resulting mesoscopic field theoretical model has been validated against known results obtained with minimal agent-based models. We found a threshold value for the activity, necessary to induce motion for a particle. Collective motion and vortex formations have been identified, as well as oscillatory motion, depending on the considered confinement. All these results are in agreement with the results in \cite{Grossman2008}. For larger systems, we analyze the formation of a traveling crystal if prepared from an initially disordered state. The traveling crystals emerge through a coarsening process from a multidomain texture of domains traveling collectively in different directions. For lower volume fractions we could identify giant number fluctuations. As theoretically predicted the standard deviation $\Delta N$ scales as $N^\alpha$ for active systems. The computed exponent $\alpha$ as a function of volume fraction is in agreement with experimental and simulation results obtained for light activated colloidal particles \cite{Palacci2013}.

The proposed mesoscopic field theoretical model can be extended from two to three spatial dimensions. Other possible extensions consider binary mixtures and hydrodynamic interactions, which are already considered within the phase field crystal model for passive systems, e.g. in \cite{Elder_PRE_2010} and \cite{Goddard2013,Toth2014,Praetorius2015,Heinonen2016}, respectively. Together with efficient numerical algorithms, see e.g. \cite{Chao2013,PaetoriusSIAM2015}, this provides the possibility to study emerging macroscopic phenomena in active systems with microscopic details, e.g. to validate coarse grained approaches, as considered in \cite{WittkowskietalNC2014,SpecketalPRL2014}.

  %acknowledgments
\ack
This work is funded by the European Union (ERDF) and the Free State of Saxony via the ESF project 100231947 (Young Investigatorts Group Computer Simulations for Materials Design- CoSiMa). We used computing resources provided by JSC within project HDR06.
\clearpage

\clearpage
\section{Bibliography}
\bibliography{activePfc}

\end{document}